# Anomalous nonreciprocal topological networks: stronger than Chern insulators

Zhe Zhang[1], Pierre Delplace[2], and Romain Fleury[1]*

[1]*Laboratory of Wave Engineering, School of Electrical Engineering, EPFL, Station 11, 1015 Lausanne, Switzerland*

[2]*Univ Lyon, ENS de Lyon, Univ Claude Bernard, CNRS, Laboratoire de Physique, F-69342 Lyon, France*

*To whom correspondence should be addressed. Email: romain.fleury@epfl.ch

**Robustness against disorder and defects is a pivotal advantage of topological systems**[1]**, manifested by absence of electronic backscattering in the quantum Hall**[2] **and spin-Hall effects**[3]**, and unidirectional waveguiding in their classical analogs**[4,5]**. Two-dimensional (2D) topological insulators**[4–13]**, in particular, provide unprecedented opportunities in a variety of fields due to their compact planar geometries compatible with the fabrication technologies used in modern electronics and photonics. Among all 2D topological phases, Chern insulators**[14–24] **are to date the most reliable designs due to the genuine backscattering immunity of their non-reciprocal edge modes, brought via time-reversal symmetry breaking. Yet, such resistance to fabrication tolerances is limited to fluctuations of the same order of magnitude as their band gap, limiting their resilience to small perturbations only. Here, we tackle this vexing problem by introducing the concept of anomalous non-reciprocal topological networks, that survive disorder levels with strengths arbitrarily larger than their bandgap. We explore the general conditions to obtain such unusual effect in systems made of unitary three-port scattering matrices connected by phase links, and establish the superior robustness of the anomalous edge modes over the Chern ones to phase link disorder of**

**arbitrarily large values. We confirm experimentally the exceptional resilience of the anomalous phase, and demonstrate its operation by building an ideal anomalous topological circulator despite its arbitrary shape and large number of ports. Our results pave the way to efficient, arbitrary planar energy transport on 2D substrates for wave devices with full protection against large fabrication flaws or imperfections.**

Among the unique and counterintuitive attributes of topological systems, topological robustness[1] against disorder and flaws is undoubtedly one of the most remarkable. This property shows ground-breaking application potential by relaxing the drastic constraints imparted by fabrication tolerances, and provide a way to seamlessly route energy and information in a wide variety of two-dimensional (2D) platforms[4–27], from quantum electronics[23] to classical photonic[4,5] and phononic devices[26,27]. Topological edge states were found in systems with broken time-reversal symmetry, such as Chern insulators[14], and then extended to time reversal invariant schemes, including the Z2[3] and other symmetry-protected schemes[29], simultaneously stimulating their classical analogues[6,10,17]. So far, Chern topological edge modes[14–24] undeniably represent the most reliable solution for point-to-point energy guiding, as they provide truly unidirectional, backscattering-immune wave transport at their boundaries. They were reported in nonreciprocal artificial wave media, such as externally-biased magneto-photonic crystals[16] or mechanical systems[13] with moving[17,19,20] or time-dependent[8,24] elements. Albeit protected from the presence of local defects by the Chern number, the edge modes cannot survive the presence of distributed disorder of sufficiently strong magnitude[1,4,5,14], especially when the average amplitude of frequency fluctuations gets larger than the band gap size. This behaviour inherently confines the topological protection of Chern phases to small distributed disorder levels.

Here, we theoretically and experimentally demonstrate a new anomalous nonreciprocal topological phase that is quantitatively stronger than the Chern phase, surviving parametric fluctuations that are arbitrarily larger than the band gap size. We find such anomalous topology in unitary scattering networks made of interconnected nonreciprocal resonant scatterers coupled by non-directed phase links. We study topological phase transitions between the anomalous and Chern phases, comparing quantitatively the robustness of their unidirectional edge modes to phase link disorder by statistical averaging over many disorder realizations. Our experiments at microwave frequencies confirm the extraordinary resilience of the anomalous phase over the Chern one. We apply our findings to the design of ideally robust circulators with arbitrarily located ports and irregular shapes, and demonstrate full non-reciprocal circulation in a 6-port prototype.

Consider the nonreciprocal unitary scattering network of Fig. 1a, which consists of general three-port nonreciprocal scatterers connected by bidirectional links in a honeycomb periodic structure. The scattering elements exhibit three-fold ($C_3$) rotational symmetry, while the links impart a phase delay of $\varphi$, as represented in the zoom-in view of the unit cell (Fig. 1b). The scattering process is described by a unitary 3x3 asymmetric scattering matrix $S_0$, whose general parametrization involves only two angles $\xi$ and $\eta$, in the interval ($-\pi/2$, $\pi/2$) (see Methods). The wave propagation in the infinite network can be described by a Bloch eigenproblem, which takes into account the scattering at the nodes, described by the 6x6 unitary matrix $S(\boldsymbol{k})$, and also involves the bidirectional phase delay $\varphi$ induced by the links:

$$S(\boldsymbol{k}) \, |c(\boldsymbol{k})> = e^{-i\varphi} |c(\boldsymbol{k})>. \tag{1}$$

So far, topological unitary scattering wave networks[6,30–34] have only been implemented in *reciprocal* systems[7,35–37] exploiting two time-reversed subspaces that are never genuinely decoupled. On the contrary, our *non-reciprocal* scattering network is formally analogous to a rigorously oriented Kagome graph (see Supplementary Information Part III), described by a unitary matrix[33] $S(\mathbf{k})$, which can be mapped[38] onto the Floquet eigenproblem of a periodically-driven lattice[39–45], with the angle variable $\varphi$ taking the role of the quasi-energy. Therefore, we can truly benefit from both advantages of non-reciprocity[46], and the potentially richer topological physics of Floquet systems[44].

We used our model to explore the parameters influencing potential topological phase transitions in the network. We found the individual reflection coefficient $R$ of the nonreciprocal scatterers, despite its complex dependency on $\xi$ and $\eta$ (see Methods), to be the main control knob for the closing of the quasi-energy band gaps. We arbitrarily set $\xi = -\eta$, which guarantees that the elements remain non-reciprocal regardless of $R$. The evolution of the bulk band structure for increasing value of $|R|$ is shown in Fig. 1c. We observe a systematic closing of two of the band gaps at $|R| = 1/3$ (denoted type 2, in red) while the others (type 1, in blue) do not change much. This suggests that topological phase transitions may be controlled by the individual scatterer reflectance.

To confirm this intuition, we probe the existence of edge modes for each of these situations by calculating the band structure of a ribbon terminated by full-reflection boundary conditions at top and bottom. As depicted in Fig. 2a, both the low and high reflection cases (respectively on the left/right panels) exhibit chiral edge modes located at the walls either at the top (red line) or bottom (blue line), with profiles represented in Fig. 2b. The main difference is that the low reflection case has edge modes in every quasi-energy band gap, whereas at high reflection, they are found in every other band gaps. This low-$|R|$ behaviour is the hallmark of anomalous Floquet insulators[33,35,42,45] (AFI), which possess topological edge states despite the Chern number of all surrounding bands being zero. In contrast, the high reflection case corresponds to the Chern insulator (CI). We map out in Fig. 2c the complete topological phase diagram for every possible realization of the scattering matrix $S_0$, represented by the angle parameters $\xi$ and $\eta$. The CI and AFI region are shaded in red and blue, respectively. We also plot contour lines depicting the reflectance of the individual scatterers in the same parameter space. Remarkably, the phase diagram unambiguously demonstrates the coincidence between the 1/3 reflection contours with the topological phase transition. The diagonal line $\xi = \eta$ is the locus of reciprocal scatterers, which can only yield trivial insulators, with all type 1 band gaps closed. The centre corresponds to a semi-metallic phase, with all band gaps closed, whereas the green point is the perfect circulator case with $R = 0$, for which the bulk bands are flat and the edge modes are dispersionless (see Extended Data Fig. 2). Such situation corresponds to a phase rotation symmetric point[33], which can only occur in the anomalous (or trivial) phases.

From the band structure of Fig. 2a, we can already intuitively expect the AFI edge modes to be much more robust than their CI counterparts to quasi-energy fluctuations, even much larger than the band gap size. Indeed, the AFI phase occurs in the ballistic regime, in which reflections at nodes are low, yielding relatively flat (slow) bulk bands and large band gaps. An abrupt jump of $\varphi$ within the lattice is very likely to land in a bandgap, which necessarily carries edge modes. Conversely, in the CI phase, the probability of an edge mode being destroyed by fluctuations larger than the band gap is much larger, due to the increased width of the bulk bands[33] and the occurrence of trivial band gaps. As an example of such a situation, let us consider the transport properties of edge modes in a finite nonreciprocal network with an abrupt quasi-energy jump in the middle (Fig. 3a, right). As a reference, we also include the case of a uniform sample (left). The two hexagonal-shaped networks have three input/output ports, as shown in the top row of Fig. 3a. Network 1 (N1) consists of uniformly distributed phase links $\varphi=\pi/8$, while for network 2 (N2), a quasi-energy jump is introduced by changing all phase links in the bottom part to $\pi/2$. With numerical simulations (see Methods and Supplementary Information Part I), we then compare the propagation of the anomalous and Chern edge modes, when exciting from port 1. The anomalous phase finds itself in topological band gaps at both $\varphi = \pi/8$ and $\pi/2$ (Fig. 2a, left), whereas the Chern phase possesses a nontrivial band gap only at $\varphi = \pi/8$, while the one at $\pi/2$ is trivial (Fig. 2a, right). As shown in Fig. 3a, the anomalous edge mode crosses the interface completely unperturbed. In stark contrast, the Chern edge mode is unable to transmit to port 2 in the presence of the interface, and all the energy is guided to port 3. Besides, upon excitation at port 2 (Extended Data Fig. 5a), the anomalous edge mode is able to reach port 3, but the trivial band gap of the Chern phase reflects all the power, exciting only an exponentially decaying field localized near port 2.

We validate experimentally this fundamental distinction between the anomalous and Chern phases by designing a nonreciprocal network operating at microwave frequencies. The scatterers are ferrite circulators connected with microstrip lines (see Methods). Our experimental design, which takes into account both the frequency dispersion of the scatterers and delay lines, finds itself in the anomalous and Chern phases at 4.9 and 3.6 GHz, respectively. Modification of the phase delays of the links is induced by changing the total lengths of the microstrip lines with serpentine paths. As shown in Fig. 3b, the measured field amplitude profiles confirm the exceptional resilience of the anomalous phase to the phase jump, in perfect agreement with the numerical predictions. Further evidence is provided by the change in measured scattering parameters and field maps measured upon exciting ports 2 and 3 (Extended Data Fig. 4d-e, 5 and 6).

The exceptional resilience of the anomalous phase in the interface scenarios considered above, involving two periodic networks, raises the question of its validity also for non-periodic quasi-energy perturbations. To answer quantitatively, we consider the same hexagonal-shaped network as in the left of Fig. 3a, and impose site-dependent disorder to the phase links, with fluctuations of strength $\delta$ randomly drawn with uniform probability in the interval $\pi/8+ [-\delta/2, \delta/2]$. We then numerically extract the transmission from port 1 to port 2 for 1000 realizations of disorder, and plot its magnitude versus $\delta$ in Fig. 3c. The solid lines represent the ensemble average, and the dashed lines are the first and last quartiles (Q1 and Q3). In the clean limit ($\delta$=0), both AFI and CI phases show unitary transmission, since the edge states exist in both cases and are unperturbed. We now turn on the disorder, up to the maximal possible strength, which corresponds to a full $2\pi$ rotation of the quasi-energy, much larger than the band gap size of both AFI and CI phases (roughly $\pi/4$). Upon increasing $\delta$, the average transmission in the Chern case quickly drops to low values. Remarkably, the AFI transmission shows a markedly different behaviour, remaining near 90% even when $\delta$ reaches $2\pi$ (fully random case). Such statistically stable robustness constitutes a solid evidence of the superiority of anomalous non-reciprocal topological networks, which exhibit the unique property of surviving disorder levels arbitrarily larger than their band gap size.

Finally, we demonstrate the use of anomalous phases in a practical scenario, where an anomalous non-reciprocal topological network is used to create a robust ideal 6-port circulator with arbitrary shape and port locations. We consider an anomalous network shaped like Switzerland, and place 6 ports at the locations of six boundary cities (Fig. 4a). We aim at connecting each city to its clockwise closest neighbour, with strong non-reciprocal isolation to any other city. A picture of the fabricated prototype is shown in Fig. 4b. We sequentially excite each input of this six-port nonreciprocal network, and report the measured experimental field maps in the AFI band (Fig. 4c). Despite the presence of finite fabrication tolerances, such as the inaccuracy in the surface mounting process of the elements, and shrinking effects due to the employed reflow oven method, and regardless of the tortuous shape of the border, we observe the expected clockwise non-reciprocal circulation of the energy, consistent with simulations (Extended Data Fig. 7c). Such robustness is also observed in longer-range transmission tests between ports 1 and 4 (Extended Data Fig. 7a and b). We envision that such anomalous wave platforms may be leveraged in a new generation of multiple-input multiple-output photonic devices[12], capable of reaching an unprecedented level of robustness. Since individual reflection is the sole control knob for the transition from the CI to the AFI phase, one could foresee very practical ways to reconfigure a domain wall between the two phases, by simply affecting the matching of the scatterers, i.e. without the need for flipping a magnetic field. This opens an avenue for a new generation of wave systems[47] that can provide reconfigurable point-to-point unidirectional energy guiding, with arbitrary control over the imparted phase delays and full immunity against backscattering. Finally, the exploration of the interplay between anomalous non-reciprocal networks and non-Hermitian perturbations, such as radiation losses occurring when coupling the edge mode to the free-space continuum, represent a promising future opportunity for topologically-controlled radiation patterns, such as in multiple beam antennas for 5G communications.

# Figures

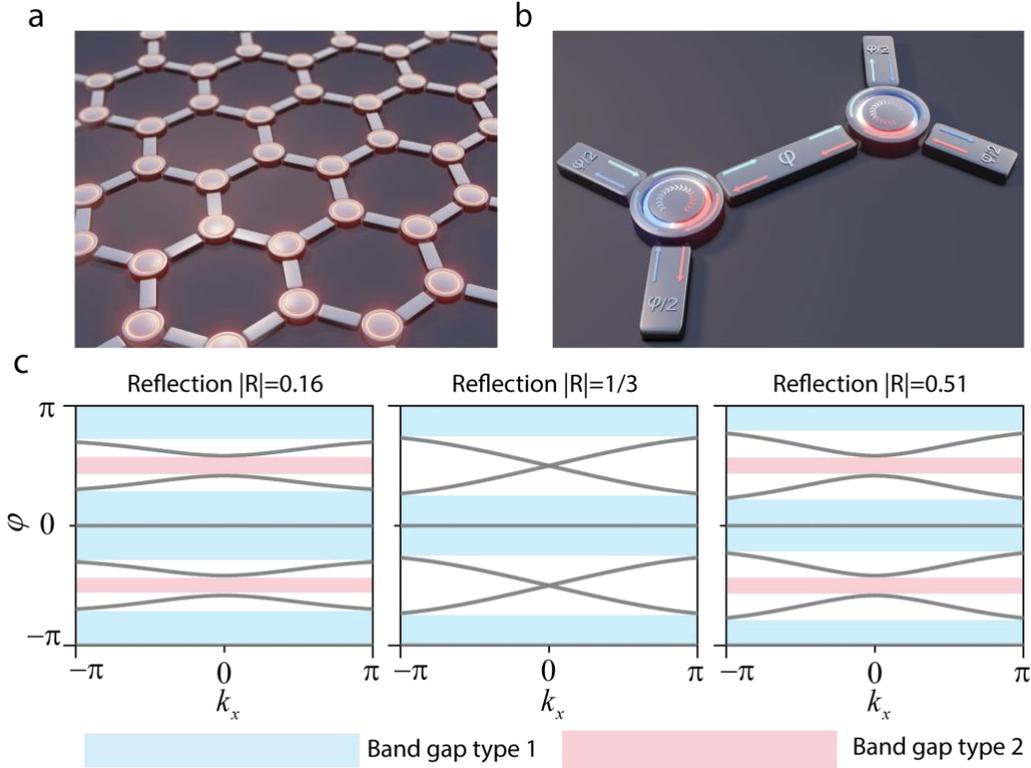

**Fig. 1: Topological non-reciprocal wave network and its bulk band structure. a,** We consider a unitary scattering network made of three-port non-reciprocal elements, described by asymmetric unitary scattering matrices. **b,** Unit cell of the honeycomb lattice, highlighting the signals entering the non-reciprocal elements, their 120° rotational symmetry, and the reciprocal phase delay $\varphi$ imparted by the links. The network is described by a unitary unit-cell scattering operator $S(\mathbf{k})$, defining a Floquet unitary mapping with quasi-energy $\varphi$. **c,** Evolution of the Floquet band structure upon increasing the level of reflection of the non-reciprocal elements from $|R| = 0.16$, (left, $\xi = -\eta = 2.5\pi/12$) to $|R| = 0.51$ (right, $\xi = -\eta = 3.5\pi/12$). At $|R| = 1/3$ (center, $\xi = -\eta = \pi/4$), the type 2 band gap closes, symptomatic of a topological phase transition.

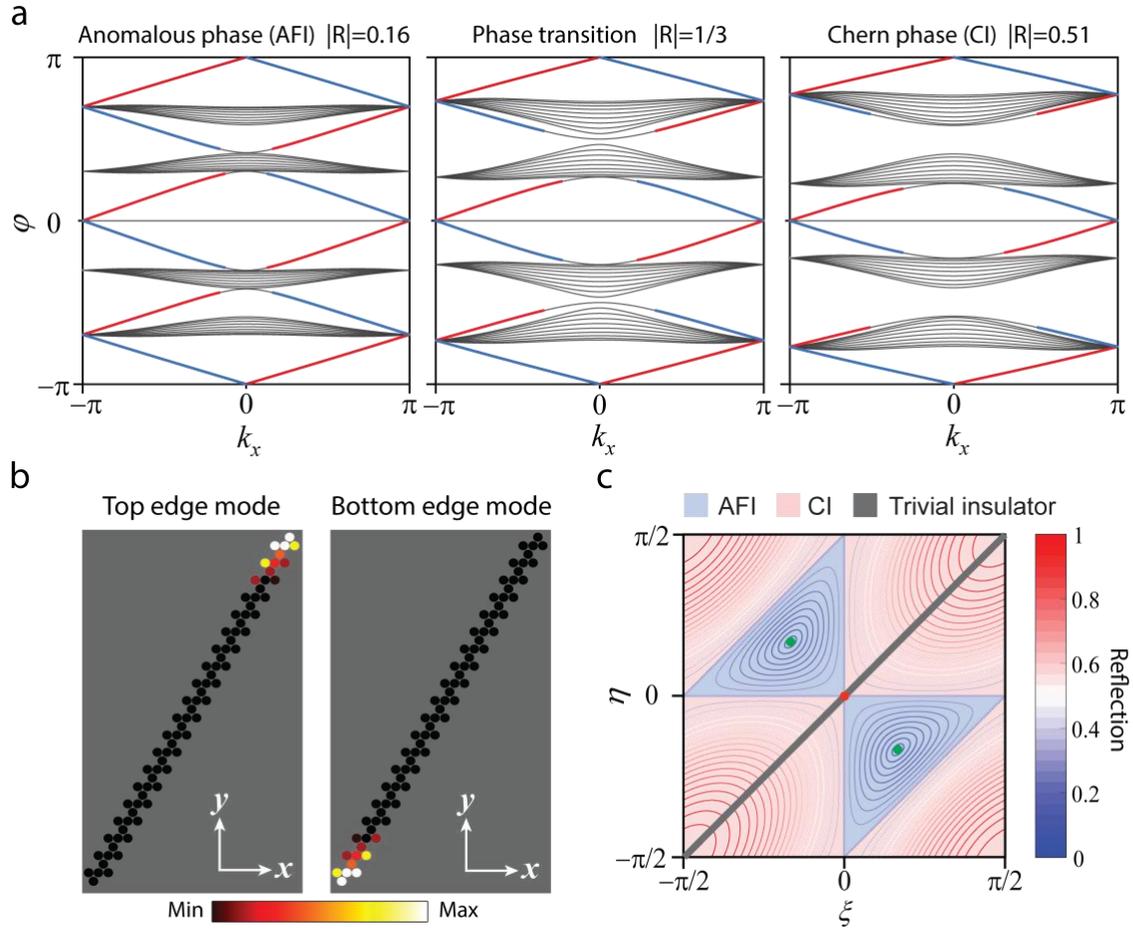

**Fig. 2: Anomalous and Chern topological phases in non-reciprocal wave networks. a,** Band structure of a supercell with periodic boundary conditions along $x$ and unitary reflection at the top and bottom. The parameters are the same as in Fig. 1c. The low reflection case is the anomalous topological phase, which features an edge mode in every quasi-energy gap. Conversely, the high reflection case supports edge modes only inside the type 1 band gaps, consistent with the Chern insulator phase. **b**, Examples of topological edge modes profiles corresponding to localization at the top and bottom of the ribbon, respectively associated with the blue and red lines in panel a. **c,** Topological phase diagram in the $(\xi, \eta)$ plane. The blue-shaded area corresponds to the anomalous phase, and the red one to the Chern phase. The lines represent the iso-reflection contours of the individual scatterers, demonstrating the coincidence between the topological phase transition and the $|R| = 1/3$ contour. On the thick gray diagonal, the scatterers are reciprocal and the type 1 band gaps close. At the center red point, all band gaps close. The green point represents the perfect circulator case.

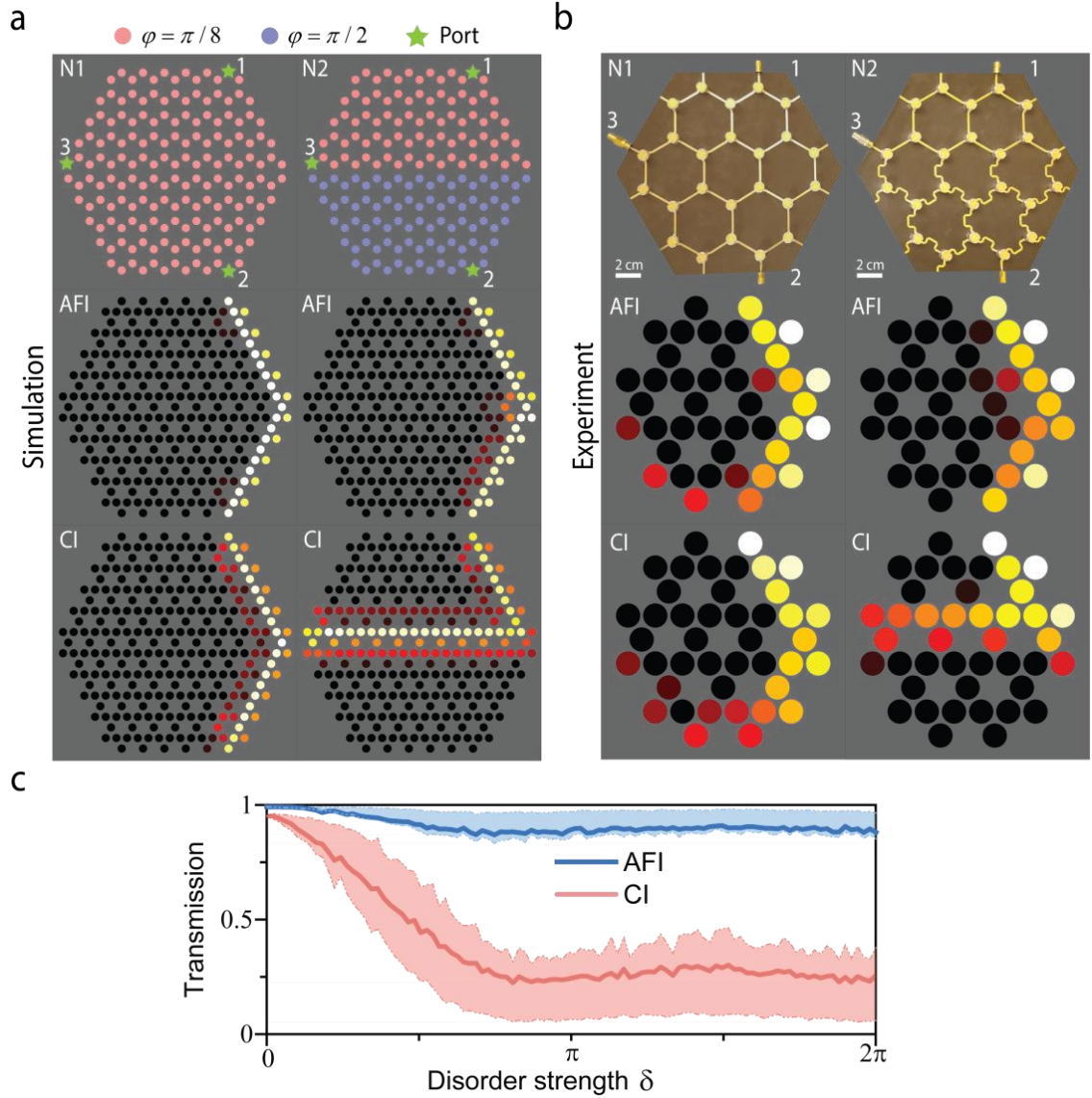

**Fig. 3: Superior robustness of anomalous nonreciprocal topological edge states. a,** Numerical simulation of the steady-state energy propagation in finite non-reciprocal networks with different phase-link distributions. The signal is incident from port 1. The parameters used to generate the anomalous (center) and Chern (bottom) phases are the same as in Fig. 1 and 2. On the left, the phase-link distribution is uniform, with $\varphi = \pi/8$, and the energy can be transmitted to port 2 in both the anomalous and Chern phases. On the right, we introduce an interface and abruptly change the value of $\varphi$ to $\pi/2$ for the bottom part. Only the anomalous phase is robust to this change, and keep transmitting to port 2. In the Chern phase, the edge mode travels along the interface and reach port 3. **b,** Experimental validation at microwaves in a network made of ferrite circulators. **c,** Transmission between ports 1 and 2 in a disordered system with randomly-generated phase delays. The phases are uniformly drawn in an interval $[-\delta/2, \delta/2]$ around $\varphi = \pi/8$. Solid lines represent the value of transmission averaged over 1000 realizations of disorder, and the dashed lines are the first and last quartiles (Q1 and Q3). The anomalous edge mode can survive disorder strengths up to a full $2\pi$ rotation.

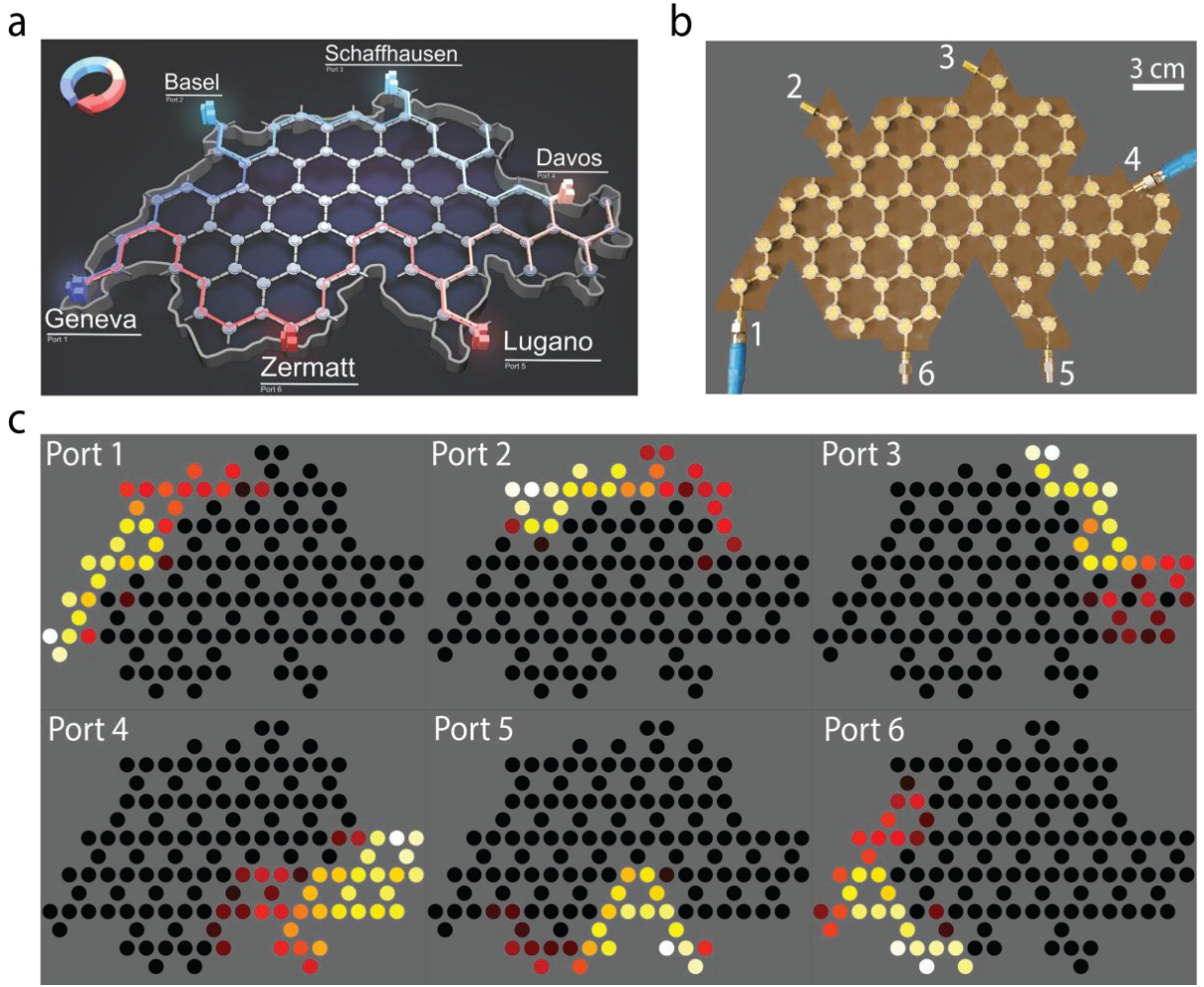

**Fig. 4: Experimental demonstration of an anomalous topological circulator with arbitrary shape and number of ports. a,** We consider a network shaped like the map of Switzerland, and placed six ports on the external boundary at six city locations. **b,** Photograph of the prototype. **c,** Experimental field maps upon sequential excitation of this 6-port system. The network behaves as a six-port circulator despite its irregular shape, the random port locations, and the high number of ports.

# Methods

**Theoretical modelling**

We first show how to parameterize a 3-port nonreciprocal element and obtain its scattering-wave model. The 3-port nonreciprocal elements, shown in Fig. 1a and b of the main text, are modelled as lossless 3-port devices with C3 symmetry, whose non-reciprocity results from Zeeman-like splitting coming from an external time-odd bias[13,48,49]. Using temporal coupled mode theory[48], their scattering is represented by an asymmetrical scattering matrix $S_0$, given by

$$S_0 = -I + \kappa^2 \begin{bmatrix} \sqrt{\gamma_+} & \sqrt{\gamma_-} \\ e^{i\theta}\sqrt{\gamma_+} & e^{-i\theta}\sqrt{\gamma_-} \\ e^{-i\theta}\sqrt{\gamma_+} & e^{i\theta}\sqrt{\gamma_-} \end{bmatrix} \begin{bmatrix} \frac{i}{(\omega-\omega_+ + i\gamma_+)} & 0 \\ 0 & \frac{i}{(\omega-\omega_- + i\gamma_-)} \end{bmatrix} \begin{bmatrix} \sqrt{\gamma_+} & e^{-i\theta}\sqrt{\gamma_+} & e^{i\theta}\sqrt{\gamma_+} \\ \sqrt{\gamma_-} & e^{i\theta}\sqrt{\gamma_-} & e^{-i\theta}\sqrt{\gamma_-} \end{bmatrix}, \quad (2)$$

where $I$ is a 3 by 3 identity matrix, $\kappa = \sqrt{2/3}$, and $\theta = 2\pi/3$, while $\omega_+$ and $\omega_-$ represent the eigenvalues of the right-handed and left-handed eigenmodes of the cavity, respectively. Zeeman-like splitting assumes that these two eigenvalues originate from a linear lifting of the two degenerate modes of the cavity by the external bias, originally at resonant frequency $\omega_0$. $\gamma_+$ and $\gamma_-$ are the inverse of their corresponding decay times to the three ports connected to the outer links, namely waveguides. Without loss, the scattering matrix $S_0$ is unitary. Due to C3 symmetry, we obtain the equality $\gamma_+ = \gamma_- = \gamma$. In the above expression, despite being an important parameter for the cavity, $\gamma$ just scales $\omega_+$ and $\omega_-$. Therefore, to show a general parameterization, we transform $\omega_+$ and $\omega_-$ into two angle variables $\xi$ and $\eta$, by standard normalizations and arctangent transformations:

$$\begin{cases} \xi = \text{atan}\,(\dfrac{\omega - \omega_+}{\gamma}) \\ \eta = \text{atan}\,(\dfrac{\omega - \omega_-}{\gamma}) \end{cases}. \tag{3}$$

$\xi$ and $\eta$ are defined in $(-\pi/2, \pi/2)$ with a periodicity of $\pi$, and characterize the deviation of the angular frequency $\omega$ from the right- and left-handed eigenvalues $\omega_+$ and $\omega_-$, respectively. Specifically, the condition $\xi = \eta$ corresponds to the reciprocal case, with $\omega_+ = \omega_-$, while $\xi = -\eta$ represents the operation at the resonant frequency $\omega = \omega_0$, with the largest non-reciprocity[48]. With the parameters $(\xi, \eta)$, the scattering matrix in Eq. (2) is rewritten as:

$$S_0 = -I + \frac{2}{3}\begin{bmatrix} 1 & 1 \\ e^{i\frac{2\pi}{3}} & e^{-i\frac{2\pi}{3}} \\ e^{-i\frac{2\pi}{3}} & e^{i\frac{2\pi}{3}} \end{bmatrix}\begin{bmatrix} \cos\xi \cdot e^{i\xi} & 0 \\ 0 & \cos\eta \cdot e^{i\eta} \end{bmatrix}\begin{bmatrix} 1 & e^{-i\frac{2\pi}{3}} & e^{i\frac{2\pi}{3}} \\ 1 & e^{i\frac{2\pi}{3}} & e^{-i\frac{2\pi}{3}} \end{bmatrix}. \tag{4}$$

As shown in Eq. (4), the individual reflection coefficient $R$ of the nonreciprocal scatterers is a function of $\xi$ and $\eta$, expressed as $R = \left|-1 + \frac{2}{3}\cos\xi\, e^{i\xi} + \frac{2}{3}\cos\eta\, e^{i\eta}\right|$.

Having established the general scattering-wave model of C3 symmetric unitary 3-port scatterers, we now derive the eigenequation Eq. (1) in the main text, which determines the bulk band structures. We give the detailed schematic of the unit cell of the periodic nonreciprocal network and signal labeling convention in Extended Data Fig. 1. Here, $S_A$ and $S_B$ are the scattering matrices of the two nonreciprocal elements A and B, respectively, governed by the parameterization Eq. (4). In the following derivations, elements A and B are the same, while the total phase delay between two nonreciprocal elements is $\varphi$. For a full description, in a unit cell of a honeycomb lattice with index $n$, the scattering waves are labeled and arranged into three vectors: $|a_n>$, $|b_n>$, and $|c_n>$, which all contain six complex wave amplitudes and represent scattering waves amplitudes propagating out, in and between the nonreciprocal elements, respectively. Based on the scattering matrices of nonreciprocal element A and B, we relate $|a_n>$ with $|c_n>$ by a unitary matrix $S_{elements}$, expressed as

$$S_{elements}|c_n\rangle \equiv P_0 \begin{bmatrix} & S_A \\ S_B & \end{bmatrix}|c_n\rangle = |a_n\rangle, \tag{5}$$

in which $P_0 = U_{12}U_{13}U_{45}U_{56}I_6$, $I_6$ is a 6 by 6 identity matrix, and $U_{ij}$ is a special permutation matrix which leads the interchange between $i$ th row and $j$ th rows of a matrix $A$ when $A$ is pre-multiplied by $U_{ij}$. Therefore, $P_0$ is also a permutation matrix, hence unitary. Here, $S_{elements}$ is a function of $(\xi, \eta)$, but not of the lattice index $n$, due to periodicity. With Bloch theory, Eq. (S4) can be written in momentum space as $S_{elements}|c(\mathbf{k})>= |a(\mathbf{k})>$.

To form the Bloch eigenproblem, we use the phase delay relation induced by the links between the nonreciprocal elements, and form a relation between $|a(\mathbf{k})>$ and $|c(\mathbf{k})>$,

$$|a(\mathbf{k})\rangle = e^{-i\varphi}\Lambda(\mathbf{k})|c(\mathbf{k})\rangle, \tag{6}$$

where $\Lambda(\mathbf{k}) = diag(e^{i\mathbf{k}\cdot\mathbf{a_2}}, e^{i\mathbf{k}\cdot\mathbf{a_1}}, 1, 1, e^{-i\mathbf{k}\cdot\mathbf{a_2}}, e^{-i\mathbf{k}\cdot\mathbf{a_1}})$ is also unitary. Substituting Eq. (6) into Eq. (5), we finally arrive the eigenequation Eq. (1) of the main text

$$S(\mathbf{k})|c(\mathbf{k})\rangle = e^{-i\varphi}|c(\mathbf{k})\rangle, \tag{7}$$

where $S(\mathbf{k}) = \Lambda^{-1}(\mathbf{k})S_{elements}$ is unitary, due to the unitarity of $S_{elements}$ and $\Lambda(\mathbf{k})$, which guarantees real-valued $\varphi(\mathbf{k})$, as $e^{-i\varphi(\mathbf{k})}$ is the eigenvalue of $S(\mathbf{k})$.

**Simulations**

The simulation method of arbitrary finite nonreciprocal honeycomb networks is based on the scattering matrix method. For a finite nonreciprocal network with $N_r$ input/output ports, once we have the information of the scattering matrix of each nonreciprocal element and the distribution of the phase delays of the links, this method can provide (i) the scattering matrix $S_{Nr}$ regarding of the

$N_r$ port system, and the field map across the network knowing the excitations at the $N_r$ ports (see details in Supplementary Information Part I).

We exemplify this method by calculating the transmission between Geneva and Davos through the Switzerland-shaped network (the network used in Fig. 4 of the main text) as a function of $\varphi$, and compare the transmission results with the ribbon band structures, as shown in Extended Data Fig. 3. We assume a uniform distribution for the phase delay $\varphi$ and the same nonreciprocal elements (in anomalous or Chern phase) in the Switzerland-shaped network. When both anomalous and Chern phases fall in a topological band gap, the transmission is near unity. When both phases fall in a bulk band, the transmission undergoes sharp variations with $\varphi$, depending on excited bulk mode. Only the Chern phase exhibit bands of blocked transmission, due the trivial band gaps.

**Design of nonreciprocal topological networks at microwave-frequency**

The nonreciprocal networks are designed and fabricated on 0.508 mm thick Rogers RT/duroid 5880 substrate (dielectric loss tan δ = 0.0009 at 10 GHz) with 35 μm thick copper on each side. Here, the nonreciprocal element is a surface mount microwave circulator UIYSC9B55T6 from UIY company, designed from a Y-shaped strip line on a printed circuit board[49]. The three ports are placed 120° apart from each other such that they are iso-spaced. The printed circuit board is sandwiched between two pieces of ferrite. Without magnetic fields, the Y-junction strip line supports two degenerate modes at $\omega_0$: right and left-handed. To bias it, two strong magnets are fixed outside, providing a strong magnetic field, normal to the printed circuit board and polarizing the ferrite, therefore lifting the initial degeneracy, with chiral modes at $\omega_+$ and $\omega_-$. In our experiment, we first measure an individual circulator and retrieve its scattering matrix $S_0$. The measured reflection of an individual circulator is shown in Extended Data Fig. 4a, and sets the frequency bands for CI and AFI operations.

Microstrip lines serve as phase delay links, with a width of 1.65 mm, corresponding to a standard 50 Ohms characteristic impedance. The phase delay $\varphi$ induced by a microstrip line with a length of $L$ operating at frequency $f$ is expressed as $\varphi = 2\pi \frac{f}{c} \sqrt{\varepsilon_{eff}} L$, where $\sqrt{\varepsilon_{eff}}$ is the effective permittivity of the microstrip line, and can be determined by an empirical formula[50]. Taking into account the frequency dispersion of the lines and circulators, we construct a more practical topological band gap map, shown in Extended Data Fig. 4b, as a function of the effective length of the microstrip lines $L$ and the operating frequency $f$. With the aid of the map, we select $L1 = 26.5$ mm and $L2 = 37.5$ mm, which produce the conditions $\varphi = \pi/8$ and $\varphi = \pi/2$ in the simulations (Fig. 3a, Extended Data Fig. 5a and Fig. 6a), respectively. As exhibited in Extended Data Fig. 4c, the fabricated networks show the microstrip lines of L1 (blue dashed region) and L2 (red dashed region).

**Measurement setup**

The scattering parameters and field maps of three fabricated networks (network 1, network 2, and the Switzerland-shaped network) are measured by a vector network analyser (VNA) R&S ZNB20, as demonstrated in Extended Data Fig. 8. For the scattering parameter measurements (Extended Data Fig. 4), as the networks are multiport, we connect the two ports of the VNA to two ports of the measured network, with the other network ports perfectly matched with 50-ohm terminations (no reflection). For the longer-range transport measurement shown in Extended Data Fig. 7, we connect ports 1 and 4 to the two VNA ports, while letting ports 2 and 4 open (full reflection) and perfectly matching ports 5 and 6. For the field map measurements, we connect the signal input port of the measured network to VNA port 1, while perfectly matching the other ports of the network. We manually probe the field at the middle of the microstrip lines by using a coaxial probe, which is connected to VNA port 2, as shown in Extended Data Fig. 8b.

## Acknowledgements

Z.Z. and R.F. acknowledge funding from the Swiss National Science Foundation under the Eccellenza award number 181232. P.D. acknowledges support from the IDEX Lyon Breakthrough program ToRe.

## Author contributions

Z.Z. performed the numerical simulations and experiments, under the supervision of R.F. P.D. and R.F. conceived the project. All authors participated in writing and revising the manuscript.

## Competing Interests

The authors declare that they have no competing financial interests.

# Extended Data

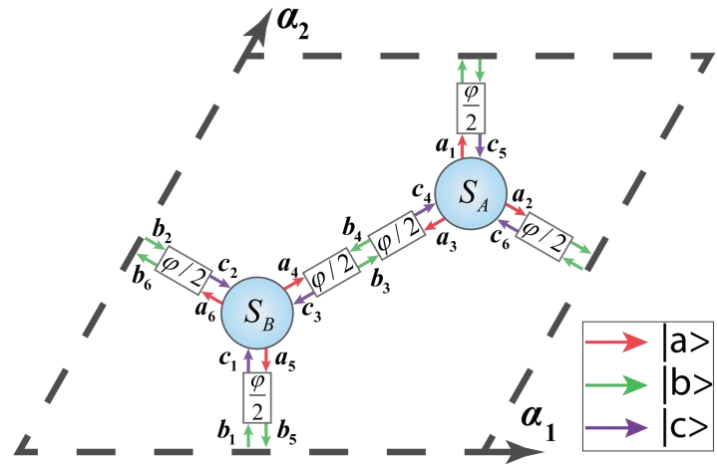

**Extended Data Fig. 1: Detailed schematic of the unit cell of the nonreciprocal network and signal labelling convention**. We define three state vectors: $|a>$, $|b>$, and $|c>$, which represent scattering waves amplitudes propagating out, in and between the nonreciprocal elements, respectively. The total phase delay between two scatterers is $\varphi$.

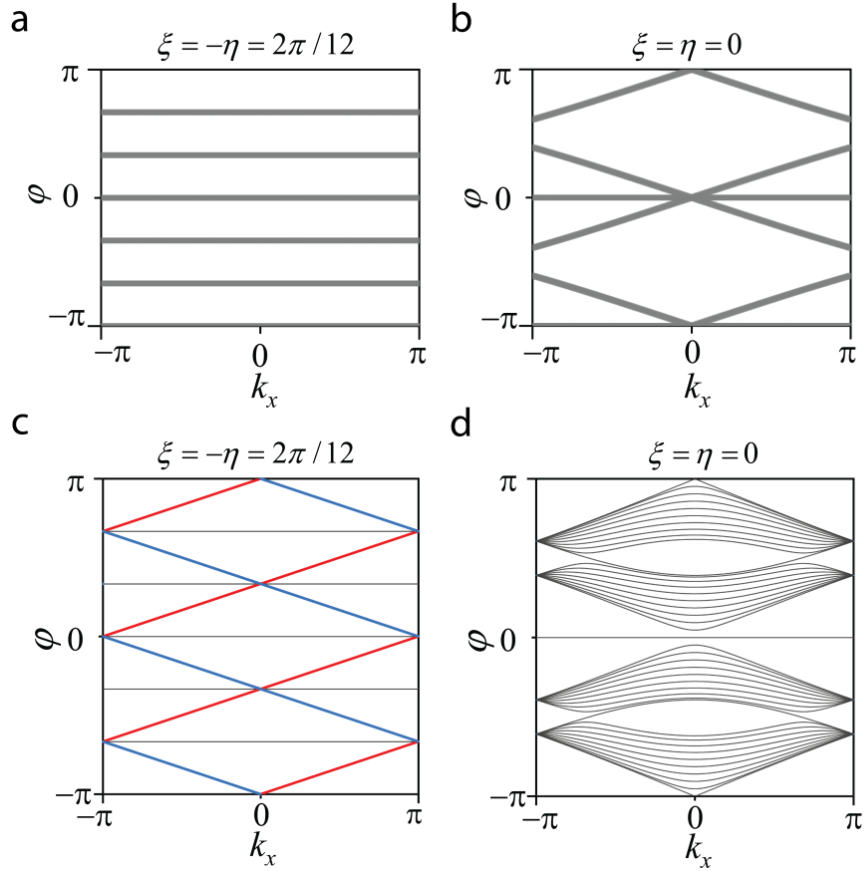

**Extended Data Fig. 2: Floquet band structures at two special points of the topological phase diagram. a, b,** Bulk band structures at the green (**a**) and center (**b**) points of the phase diagram of Fig. 2c in the main text. The green point corresponds to a phase-rotation symmetric network of perfect matched circulators, thus in AFI phase. The red center point represents a network of reciprocal resonant scatterers, with all band gaps closed. **c, d,** Ribbon band structures corresponding to panel a and b. The perfect circulator network features flat bulk band with dispersionless edge modes regardless of the value of the quasi-energy $\varphi$, which can only occur in the AFI phase.

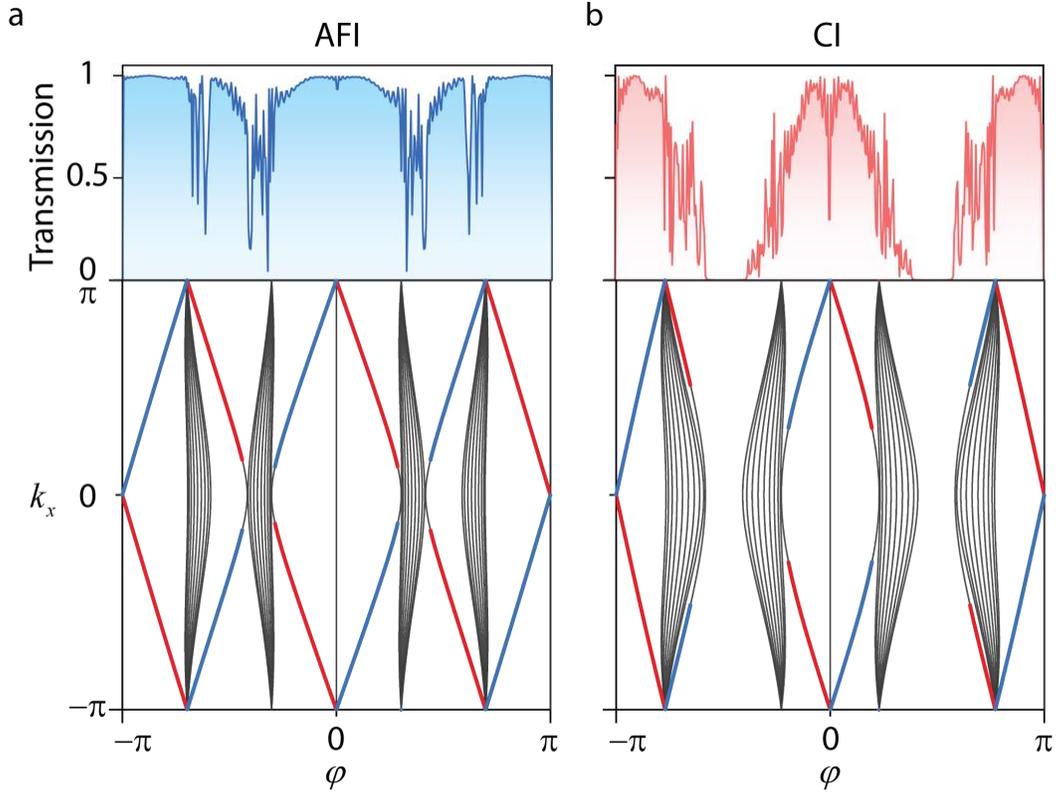

**Extended Data Fig. 3: Comparison between 2-port transport properties and the ribbon band structures.** We plot the transmission between Geneva and Davos through the Switzerland-shaped network as a function of $\varphi$, and compare it with the ribbon band structure. We assume a uniform distribution for the phase delay $\varphi$. **a,** case of the anomalous phase in Fig. 1c. **b,** case of the Chern phase shown in Fig. 1c. When $\varphi$ falls in a topological band gap, transmission is mediated by the edge modes and reach high values. Conversely, if $\varphi$ belongs to a trivial band gap, transmission is impeded. Finally, if $\varphi$ falls in a bulk band, the transmission fluctuates with $\varphi$, depending on the excited bulk modal superposition interference at the output port.

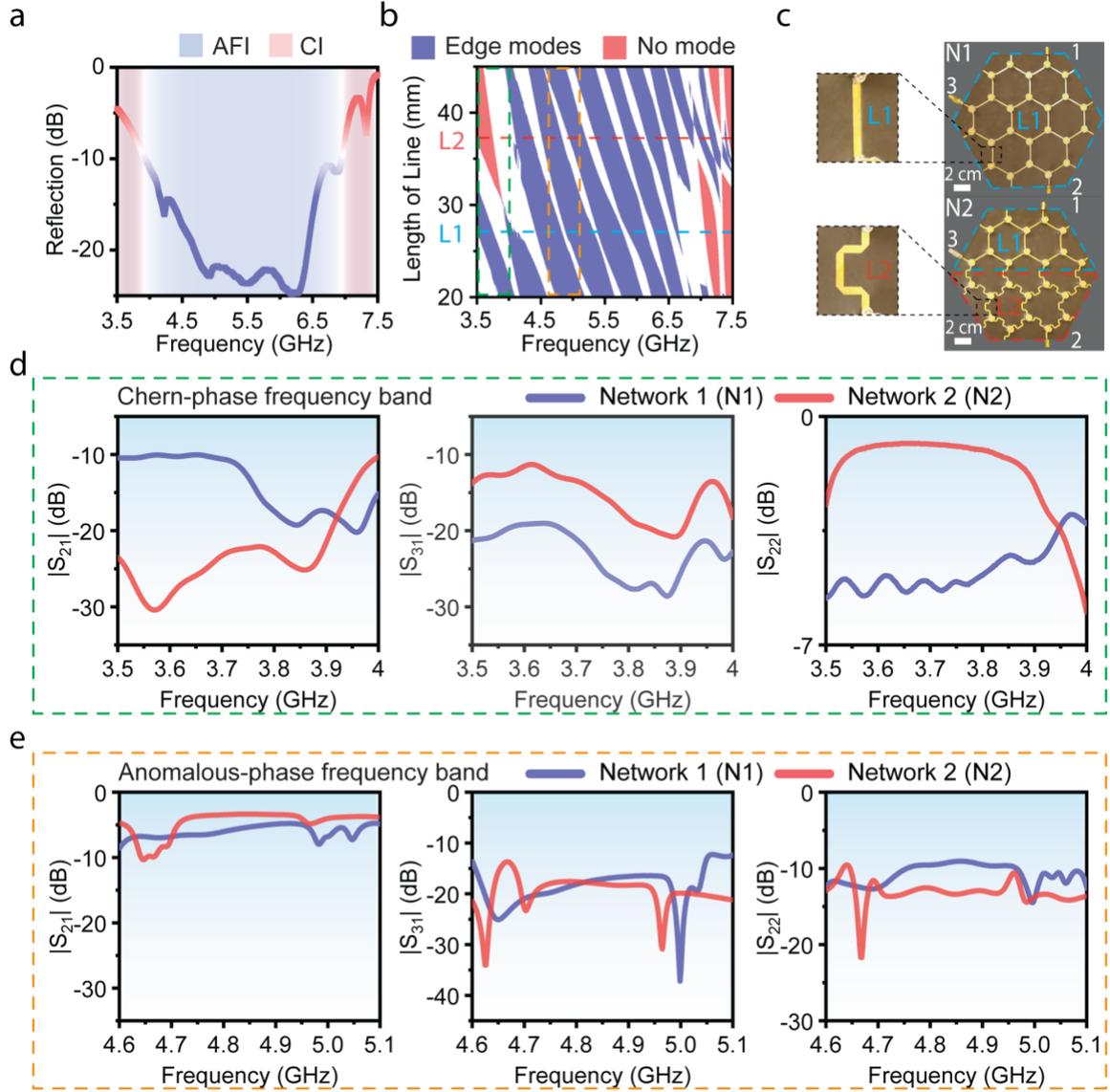

**Extended Data Fig. 4: Experimental network design and measured scattering parameters. a,** Measured reflection spectrum of an individual ferrite circulator. The blue-shaded area represents the bandwidth of the anomalous phase, corresponding to low reflection ($|R|<-9.5$ dB$=20\cdot\log_{10}(1/3)$). By contrast, the red-shaded area shows the Chern phase with high reflection ($|R|>-9.5$ dB). Topological phase transitions happen at around 3.9 GHz and 7 GHz. **b,** Topological band gap map predicted from the individual scattering data, when varying the length of the microstrip connections and the operating frequency. The blue and red regions correspond to band gaps with and without topological edge modes, respectively. The white regions represent bulk bands. **c,** Design details of the experimental networks probed in Fig. 3b of the main text. Network 1 (N1) has a uniform length distribution of microstrip lines with L=L1. For network 2 (N2), we introduce an interface and replace the bottom part with lines of different length L2. **d,** Measured amplitudes of the scattering parameters $S_{21}$ (left), $S_{31}$ (mid) and $S_{22}$ (right) in the Chern-phase frequency band (green dashed box in panel b). **e,** Measured scattering parameters in the anomalous-phase frequency band (yellow dashed box in panel b).

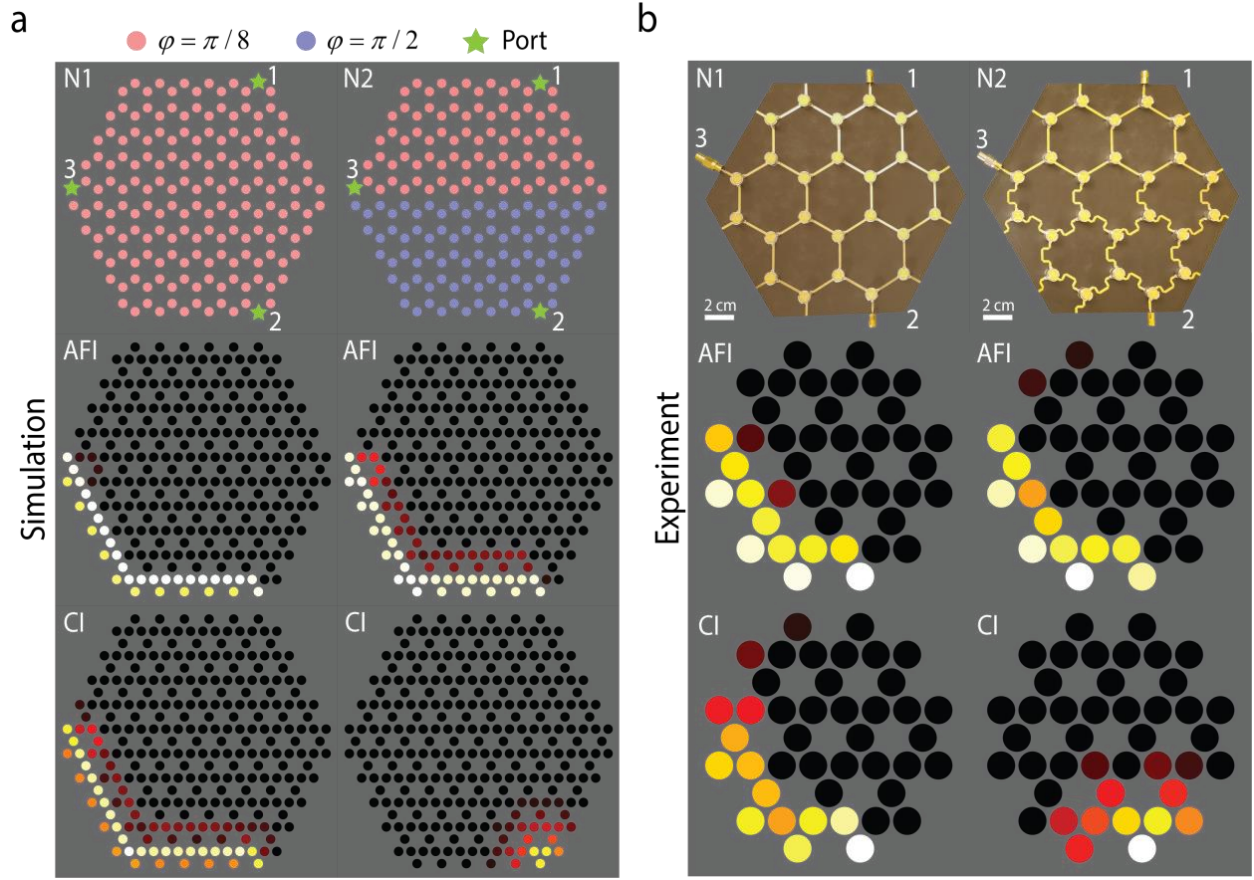

**Extended Data Fig. 5: Numerical and experimental field maps for excitation at port 2. a,** Numerical predictions for excitation at port 2 for the same system as in Fig. 3 of the main text. While the anomalous phase supports transmission to port 3 regardless of the phase link distribution, the Chern phase possesses a trivial band gap at $\varphi = \pi/2$, and reflects all the energy incident from port 2, see bottom right plot (the field distribution exhibits exponential decay). **b,** Corresponding experimental data.

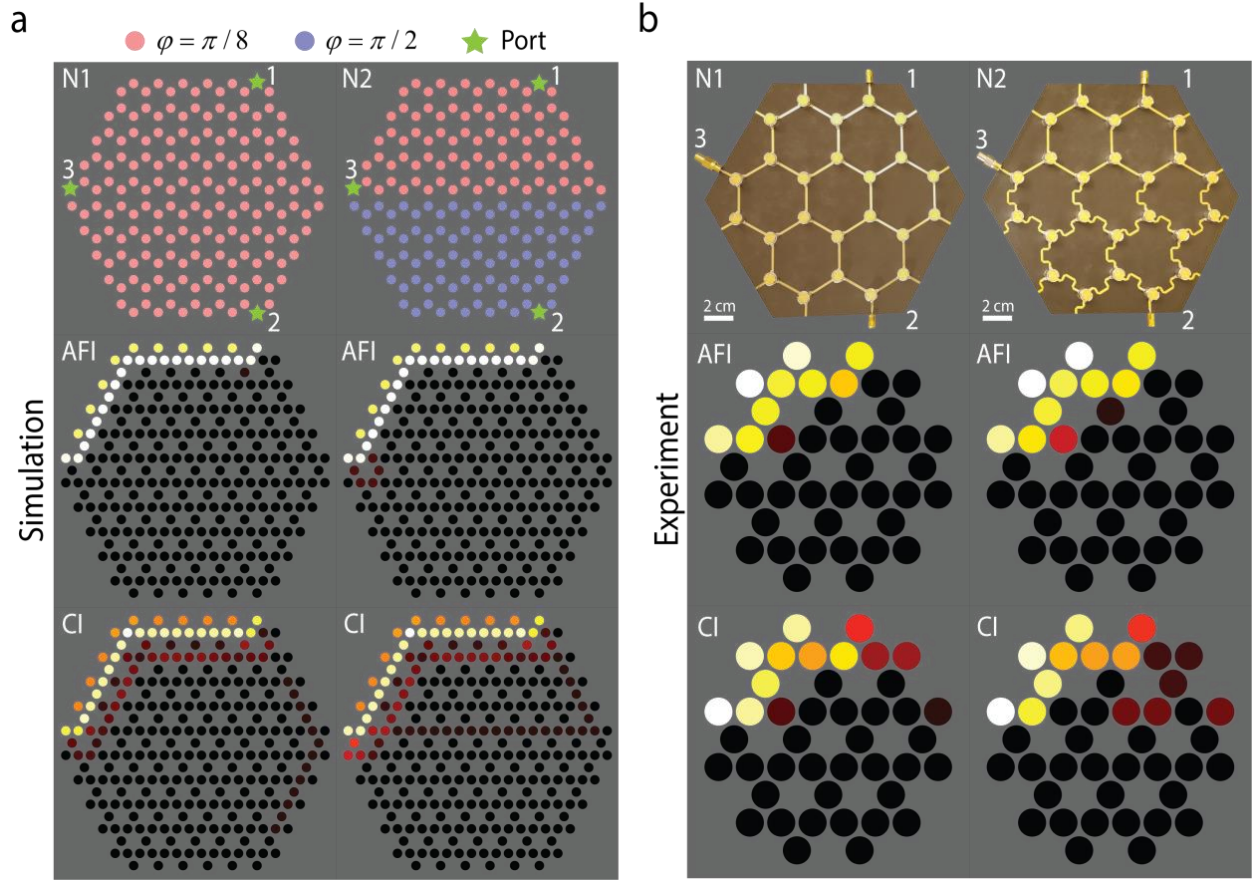

**Extended Data Fig. 6: Numerical and experimental field maps for excitation at port 3. a,** Numerical predictions for excitation at port 3 for the same system as in Fig. 3 of the main text. Both the anomalous and Chern phases fall in topological band gap at $\varphi = \pi/8$, leading to transmission to port 1. **b,** Corresponding experimental data.

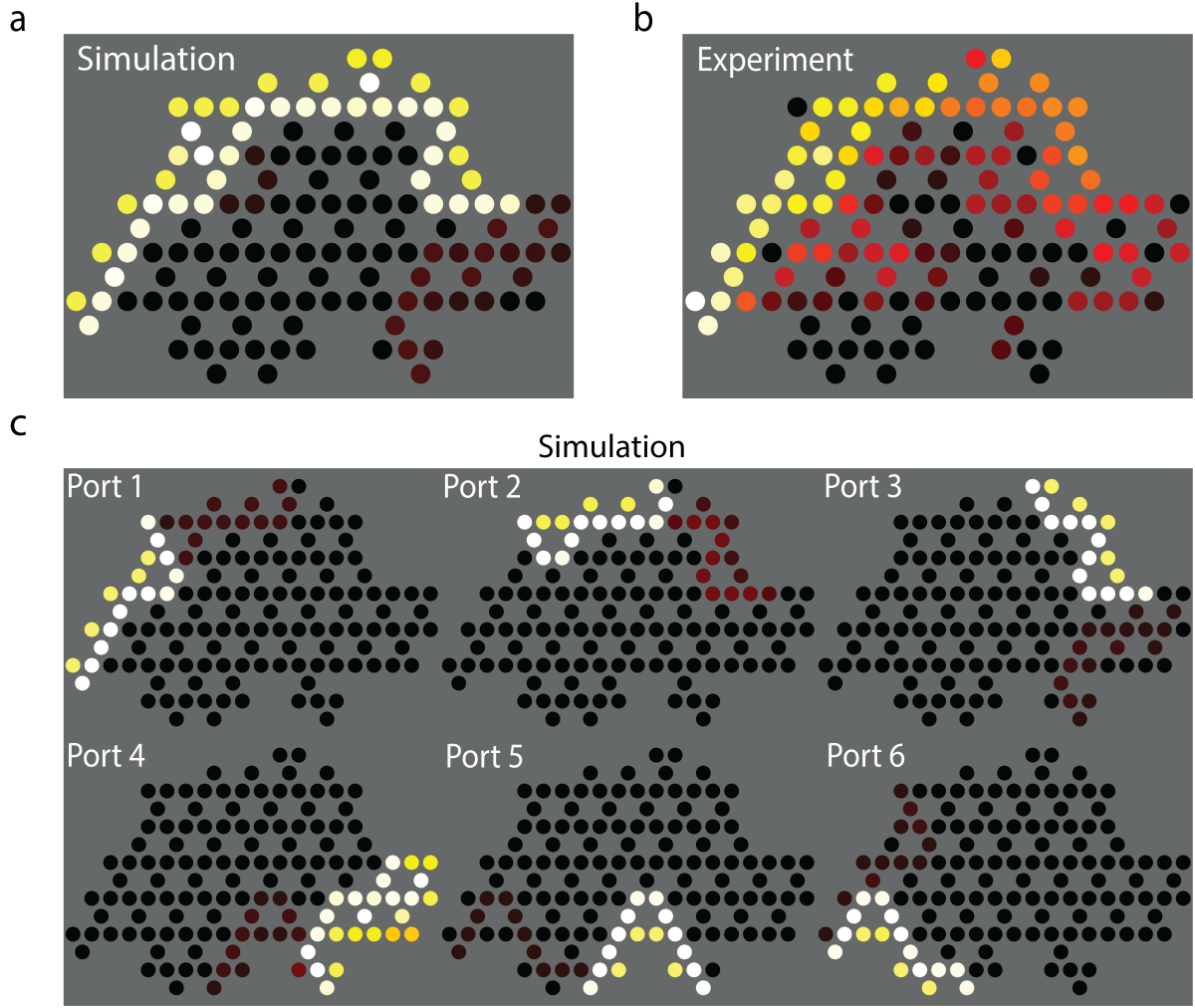

**Extended Data Fig. 7: Additional field maps for the anomalous topological Switzerland-shaped network.** We plot simulated (**a**) and experimental (**b**) transmissions from Geneva (port 1) to Davos (port 4) for the same network in Fig. 4 of the main text, leaving all other ports open. **c,** Numerical prediction corresponding to the experimental data shown in Fig. 4c of the main text.

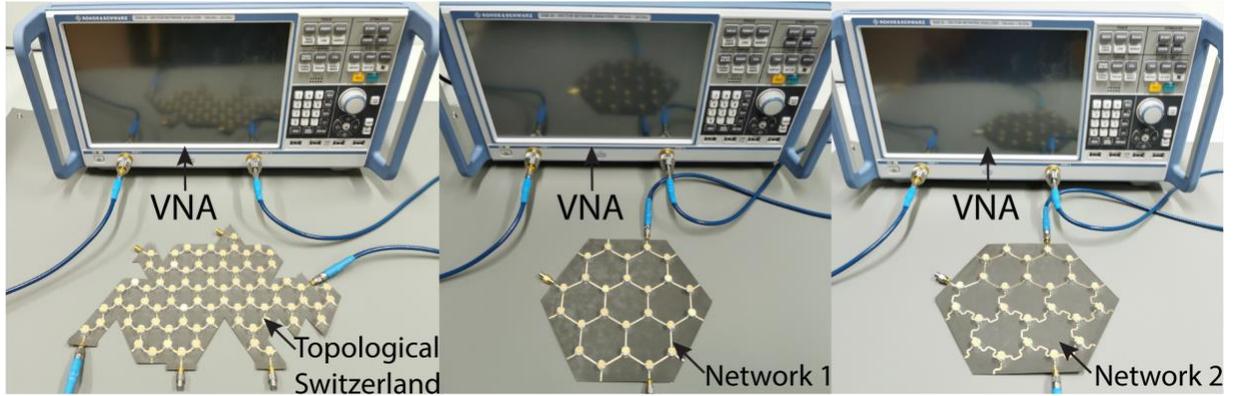

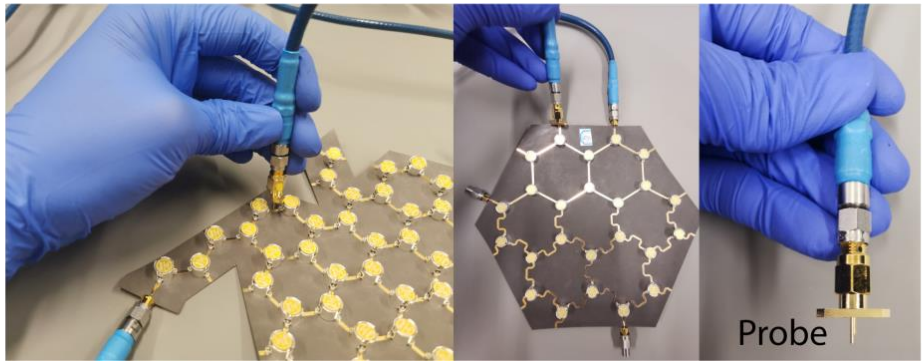

**Extended Data Fig. 8: Experimental setups for scattering parameter and field distribution measurements. a,** The setup consists of a vector network analyser (VNA) and three microwave nonreciprocal networks: the Switzerland-shaped network (left), N1 (middle), and N2 (right). **b,** Field map measurement with a coaxial probe for measuring fields on the microstrip lines.